\documentclass[12pt]{article}
\usepackage{amsmath}
\usepackage{amssymb}
\begin{document}
\vskip 2cm
\begin{center}
{\sf {\Large ${\mathcal N} = 2 $ Supersymmetric Harmonic Oscillator: Basic Brackets 
Without Canonical Conjugate Momenta 
}}

\vskip 2.0cm

{\sf N. Srinivas$^{(a)}$, A. Shukla$^{(a)}$\footnote{Present address:  Indian Institute of Science Education and 
Research Kolkata, $~~~~~~~~~~~~~~~~~~~~~~~~~~~~~~~~~~~~~~$ Mohanpur - 741 246,  
West Bengal, India},  R. P. Malik$^{(a,b)}$}

\hskip 1cm $^{(a)}$  Physics Department, Centre of Advanced Studies,\\
\hskip 1.5cm {\it Banaras Hindu University (BHU), Varanasi - 221 005,  India}\vskip .06cm
$^{(b)}$DST Centre for Interdisciplinary Mathematical Sciences,\\
\hskip 1.5cm {\it Faculty of Science, Banaras Hindu University, Varanasi - 221 005, India}\vskip .06cm
\hskip 1.5cm {\small {\sf {E-mails: seenunamani@gmail.com; ashukla038@gmail.com;  rpmalik1995@gmail.com}}}

\end{center}

\vskip 1cm

\noindent
{\bf Abstract:} 
We exploit the ideas of spin-statistics theorem, normal-ordering and the key concepts
behind the symmetry principles to derive the canonical (anti)commutators for the case of a one (0
+ 1)-dimensional (1D) $\mathcal{N} = 2$ supersymmetric (SUSY) harmonic oscillator (HO) without taking
the help of the mathematical definition of canonical conjugate momenta with respect to the bosonic
and fermionic variables of this toy model for the Hodge theory (where the continuous and discrete
symmetries of the theory provide the physical realizations of the de Rham cohomological operators
of differential geometry). In our present endeavor, it is the full set of continuous symmetries and
their corresponding generators that lead to the derivation of basic (anti)commutators amongst the
creation and annihilation operators that appear in the normal mode expansions of the dynamical
fermionic and bosonic variables of our present $\mathcal{N} = 2$ SUSY theory of a HO. These basic brackets
are in complete agreement with such kind of brackets that are derived
from the standard canonical method of quantization scheme.




\newpage

\section{Introduction}	 
The quantization of a given classical
system becomes very essential when we study this
system at an energy scale where the quantum mechanical
effects become very important. For instance, for any
arbitrary classical matter, the quantum mechanical phenomena
become very prominent when we study this system
at its atomic/nuclear scale. One of the earliest methods of quantization scheme
is the standard canonical method of quantization. 
This method of quantization invokes primarily {\it three} basic ideas. 
First and foremost, we distinguish between the fermionic and bosonic variables/fields of the theory 
by using the celebrated spin-statistics theorem (which dictates the existence  of (anti)commutators 
at the {\it quantum} level for such variables/fields). Second, we define the canonical conjugate momenta
for the above variables/fields and define the (graded)Poisson brackets at the {\it classical} level which 
are promoted to the (anti)commutators at the {\it quantum} level. Finally, if the variables/fields allow normal mode 
expansions due to their equations of motion, we express the above  (anti)commutators in terms of the creation 
and annihilation operators (of the normal mode expansions). The above cited basic (anti)commutators, at the variables/fields
level, get translated into the basic (anti)commutators amongst the creation/annihilation operators. Normal-ordering is 
required to make physical sense  out of  the physical quantities (e.g., Hamiltonian, conserved charge, etc.) 
when they are expressed in terms of the creation and annihilation operators of the normal mode expansions of the variables/fields.

In our present endeavor, we shall exploit  the ideas of spin-statistics theorem\footnote{For our present 
1D toy model of SUSY-HO, the spin-statistics theorem implies the existence of (anti)commutators
at the quantum level because we cannot define the ``spin'' (i.e. Pauli-Lubansky vector) for such kind of a 1D toy model.}   and
normal-ordering but we shall {\it not} use {\it purely} mathematical definition of the canonical conjugate
momenta in the quantization of a one (0 + 1)-dimensional (1D) SUSY harmonic oscillator 
(SUSY-HO) in terms of its creation and annihilation operators. Instead, we shall utilize the ideas of symmetry 
principles (i.e. continuous symmetries and their generators)
to obtain the basic (anti)commutators of this SUSY system (which has been proven to be a model for the Hodge theory in our
earlier work [1]). The central claim of our present investigation is the observation that the quantization of a 
class of theories  can be performed {\it without} the mathematical definition of the canonical conjugate momenta. These set of theories 
belong to the models which are tractable physical examples of Hodge theory. In this context, it is very gratifying to state that in a very recent publication [2],
we have obtained the canonical brackets for the 1D model of a rigid rotor without taking any help from the 
mathematical definition of the canonical conjugate momenta. In the standard method of canonical quantization scheme,
the latter definition plays a key role.  

At the field theoretic  level, it has been shown that the 2D {\it free} Abelian 1-form  gauge theory [3,4]
and its {\it interacting} version (where there is a coupling   between  2D Abelian gauge field with Dirac fields [5])
 are tractable field theoretic models for the Hodge theory. The common features  of all the 
above citied models for the Hodge theories  is the observation that their discrete and continuous symmetry
transformations provide the physical realizations of the de Rham cohomological operators of differential geometry.
In our recent works [6,7], we have demonstrated that the covariant canonical  quantization 
of the 2D gauge theories  can be performed without the definition  of the canonical conjugate momenta.
In fact, we have shown that the symmetries  of these theories, within the framework of 
Becchi-Rouet-Stora-Tyutin (BRST) formalism, are good enough to lead to the derivation  
of the basic (anti)commutators where inputs from the spin-statistics theorem  and normal-ordering 
are required.  The noteworthy point is the observation that the mathematical definition of the canonical conjugate  momenta 
is {\it not} required for the derivation of  basic (anti)commutators amongst the creation and annihilation operators
of such a specific class of theories.

In our present investigation, we have chosen the model of 1D SUSY-HO because it contains
 bosonic as well as fermionic
variables which require (anti)commutators for their quantization (at the quantum level). 
Moreover, as we know, the range 
and reach of physics behind the system of a harmonic oscillator is very wide as
 it encompasses in its folds the theoretical ideas
from classical mechanics to field  and string theories.  
Thus, it is a very good proposition to say something {\it novel}
about such a widely applicable system of theoretical physics. 
We establish, in our present endeavor, that there is no need of the 
mathematical definition of the canonical conjugate momenta for the 
quantization of 1D SUSY-HO as its innate continuous symmetries are good enough 
to entail upon the existence of basic canonical brackets (i.e.  (anti)commutators) 
at the level of creation and annihilation operators.

Our present investigation has been motivated by the following key and decisive factors.   First, the definition  of 
the  canonical conjugate momenta is purely  {\it mathematical}  in nature. Thus, any {\it physical }
alternative to it is a welcome sign as far as the richness of ideas in theoretical physics is concerned.
Second, it is useful to derive  the basic canonical (anti)commutators from another method than the {\it usual} canonical method 
because it would enrich the tools and techniques in the realm of theoretical physics.
Third,  to put our ideas and experiences of the 2D free as well as interacting Abelian gauge theories [5,6]
on firmer-footings, it is essential to prove the sanctity of these 
ideas in the context of other examples of Hodge theory.
Our present endeavor is an attempt in that direction. 
Finally, our method of derivation of the canonical (anti)commutators
 is based on symmetry considerations. Thus, even though  algebraically more involved, our method 
of derivation is {\it physically} more appealing as far as the quantization of our system is concerned.

The contents of our present endeavor are organized as follows. In Sec. 2, we very briefly mention about the usual 
$\mathcal{N} = 2$ SUSY symmetries of the SUSY-HO and the bosonic symmetry 
that emerges from their anticommutator.
Our Sec. 3 captures 
the usual method of canonical quantization scheme to derive the (non-)vanishing basic canonical brackets.
Sec. 4  of our present paper is devoted to the derivation of basic brackets from the 
{\it first} of the $\mathcal{N} = 2$ SUSY symmetries. In Sec. 5, we derive the (anti)commutators 
that results in from the {\it other} $\mathcal{N} = 2$ SUSY transformations.
Our Sec. 6 contains  the basic (anti)commutators 
that emerge from the bosonic symmetry transformations.  Finally, we make some concluding remarks
and point out a few future directions  in our Sec. 7.

\section{Preliminaries: continuous symmetries and conserved charges as generators }   
  
  We begin with the following Lagrangian for the 1D SUSY-HO with mass $m = 1$ and natural 
frequency $\omega$ (see, e.g. [8,1] for details)
\begin{eqnarray}
L = \frac{1}{2}\,{\dot x}^2 - \frac{1}{2}\,\omega^2\, x^2 + i\, \bar\psi\,\dot\psi 
- \omega \, \bar\psi\, \psi,
\end{eqnarray}
where $\dot x = (dx/dt), \, \dot\psi = (d\psi/dt)$ are the generalized ``velocities" for the bosonic 
and fermionic variables $x$ and $\psi$, respectively. For the bosonic variable $x$, there are two fermionic 
($\psi^2 = 0, \, {\bar\psi}^2 = 0,\; \psi\, \bar\psi + \bar\psi\, \psi =0$) variables in the 
 $\mathcal{N} = 2$ SUSY theory (at the classical level).
The latter are the superpartners of the former. In other words, as we shall see in Eq. (2) below, under the
$\mathcal{N} = 2$ SUSY symmetry transformations, the bosonic and fermionic variables  transform to one-another obeying
the basic principles of the $\mathcal{N} = 2$ SUSY theory of HO.

The above Lagrangian respects the following   infinitesimal and continuous
$\mathcal{N} = 2$ SUSY symmetries\footnote{It can be checked that the Lagrangian (1) transforms to
the total time derivative under the continuous transformations (2). Thus, the action integral remains invariant for the physically 
well-defined variables which vanish off at infinity. To be precise, the  
transformations (2) are the {\it symmetry} transformations for the action integral.} (see, e.g. [1])
\begin{eqnarray}
&&s_1 x =  \psi, \qquad s_1 \psi = 0,\qquad s_1 \bar\psi = i\,(\dot x + i\, \omega \, x), \nonumber\\
&& s_2 x =  \bar\psi, \qquad s_2 \bar\psi = 0,\qquad s_2 \psi = i\,(\dot x - i\, \omega \, x),
\end{eqnarray}
which are nilpotent  (i.e. $s_1^2 = 0, \, s_2^2 = 0$) of order two on the on-shell 
($\dot \psi + i\, \omega \,\psi = 0, \; \dot {\bar\psi} - i\, \omega\, \bar\psi = 0$). 
The above continuous symmetry transformations are generated by the following 
Noether conserved ($\dot Q = \dot{\bar Q} = 0$)
 charges ($Q$ and $\bar Q$), namely; 
\begin{eqnarray}
Q = (\dot x + i\, \omega \, x)\, \psi, \qquad\qquad  \bar Q = \bar\psi\, (\dot x - i\, \omega \, x),
\end{eqnarray}
which are {\it also} nilpotent ($Q^2 =  {\bar Q}^2 = 0$) of order two  and they are conserved. 
The latter property  can be checked by using the Euler-Lagrange (EL) equations of motion\footnote{It will
be noted that these {\it classical} EL equations of motion turn into operator form at the {\it quantum} level. Thus, 
their solutions can be expressed in terms of operators (which are nothing but the creation and 
annihilation operators [cf. (8) below]).} 
\begin{eqnarray}
&&\ddot x + \omega^2\, x = 0, \quad \dot \psi + i\, \omega \,\psi = 0, \quad
 \dot {\bar\psi} - i\, \omega\, \bar\psi = 0, \nonumber\\  
&&\ddot \psi +  \omega^2 \,\psi = 0, 
\quad \ddot {\bar\psi} +  \omega^2\, \bar\psi = 0,
\end{eqnarray}
which emerge from the Lagrangian (1) of our present SUSY theory from the least action principle when we 
demand that $\delta \, S = 0$ for the action integral $S = \int dt L$. It should be emphasized, at this stage,
that the Noether theorem does not invoke the definition of the canonical conjugate momenta. Rather, it
is derived from the action principle (where the physical system follows the trajectory that is described by the 
EL equations of motion).

The anticommutator (i.e. $s_\omega = \{s_1,\, s_2\}$) leads to the definition of a
bosonic symmetry ($s_\omega$) in the theory. Under this transformation, the variables change as: 
\begin{eqnarray}
&& s_\omega \, x= \{s_1, \, s_2\}\, x = \dot x, \qquad s_\omega \, \psi= \{s_1, \, s_2\}\, \psi = \dot \psi,
\nonumber\\ 
&& s_\omega \, \bar\psi= \{s_1, \, s_2\}\,\bar \psi = \dot {\bar\psi},
\end{eqnarray}
modulo a factor of $2i$. The above transformations demonstrate the validity and existence of $\mathcal{N} = 2$ SUSY
symmetries in a  our 1D theory because the anticommutator of two SUSY transformations
 is equivalent to a time-translation. Under the above transformations, the Lagrangian transforms ($s_\omega\, L =
 \frac{d}{dt} \, [L]$) to its own time-derivative. Thus, the generator of transformations (5)
 is nothing but the Hamiltonian of our theory\footnote{It is quite elementary to check that the 
transformations ($s_1,\, s_2,\, s_{\omega}$) satisfy a beautiful algebra: $s^2_1 = s^2_2 = 0,\, 
s_{\omega} = \{s_1,\, s_2 \},\, [s_{\omega},\, s_1] = 0,\, [s_{\omega},\, s_2] = 0$ in 
their operator form. This algebra is also mimicked by the charges ($Q,\, \bar Q,
H$) which is nothing but the ${\mathcal N} = 2$ SUSY quantum mechanical algebra $sl(1/1)$ in its simplest form.
In our earlier works [1, 16], these algebras are also shown to be 
identified with the algebra of the de Rham cohomological operators
of differential geometry.}, namely;  
\begin{eqnarray}
Q_\omega &=& \frac{1}{2}\,\dot x^2 + \frac{1}{2}\,\omega^2\, x^2 + \omega \, \bar\psi\, \psi \nonumber\\
&\equiv & \frac{{p_x}^2}{2} + \frac{\,\omega^2\, x^2}{2} + \omega \, \bar\psi\, \psi \equiv H,
\end{eqnarray}
where $p_x = \dot x$ is the momentum corresponding to the bosonic variable $x$ and $H$ 
is the canonical Hamiltonian. The latter can be {\it also} derived by the Legendre  transformations
as given below, namely;  
\begin{eqnarray}
H &=& \dot x\, \Pi_x + \dot \psi\, \Pi_\psi + \dot{\bar\psi}\, \Pi_{\bar\psi} - L \nonumber\\
&\equiv & \frac{{p_x}^2}{2} + \frac{\,\omega^2\, x^2}{2} + \omega \, \bar\psi\, \psi,
\end{eqnarray}
where $\Pi_x = p_x = \dot x, \, \Pi_\psi = -i\,\bar\psi, \, \Pi_{\bar\psi} = 0$ are the canonical conjugate
momenta\footnote{Note that $\Pi_{\bar\psi} = 0$ is {\it not} a primary constraint on the theory
because the third term in the Lagrangian (1) can be symmetrized such that $\Pi_{\bar\psi} \ne 0$.}.
In the above computation, 
we have used the idea of left-derivative with respect to the fermionic variables 
(i.e. $\Pi_\psi = \partial L/\partial \dot \psi, \,\Pi_{\bar\psi} = {\partial L}/{\partial \dot {\bar\psi}} $).
This is the reason that there is a negative sign in $\Pi_\psi = - i\, \bar\psi$.

\section{Basic brackets: standard canonical method}

We begin with the following normal mode expansions for the variable $x(t), \psi(t)$ and $\bar\psi(t)$ 
of our present theory of 1D SUSY-HO (see, e.g. [9]): 
\begin{eqnarray}
&& x(t) = \frac{1}{\sqrt{2\omega}}\, [a \,e^{- i \,\omega \,t} + a^{\dagger} \,e^{+ i\, \omega\, t}],\nonumber\\
&& \psi (t) = b \,e^{- i\, \omega\, t}, \qquad \bar{\psi}(t)= b^{\dagger} \, e^{+i \,\omega \,t},
\end{eqnarray}
which satisfy the EL equations of motion (4). It is the validity of the equations of motion ($\dot \psi + i \omega \psi = 0,
\dot {\bar \psi} - i \omega \bar \psi = 0$) which forces us to choose the solutions for $\psi$ and $\bar \psi$ as given in (8).
In the above, the time-independent dagger and non-dagger operators are the creation and annihilation operators. 
The following standard canonical (anti)commutators:
\begin{eqnarray}
&& [x, \, x] = [\Pi_x, \, \Pi_x] = 0, \nonumber\\
&&  [x, \, \Pi_x] = i \,\equiv \,  [x,\, p_x], \; \Pi_x = p_x,\nonumber\\ 
&&  \{\psi, \,\psi \} = 0, \qquad \{\bar \psi,\, \bar \psi \} = 0,  \nonumber\\
 && \{ \psi,\,  \Pi_\psi \} = i \, \Longrightarrow  \, \{ \psi,\, \bar\psi \} =-1,
\end{eqnarray}
can be expressed in terms  of the creation and annihilation operators as
\begin{eqnarray}
&& [a, \,a] = [a^{\dagger}, \,a^{\dagger}] = 0 ,\quad  [a, \, a^{\dagger}] = 1,\nonumber\\
&& \{b, \, b\} = \{b^{\dagger},\, b^{\dagger} \} = 0  \quad
\{b,b^{\dagger} \} =-1, 
\end{eqnarray}
if we exploit the mode expansions given in (8) and use them in the canonical brackets (9).
In other words, the basic (anti)commutators of (9) and (10) are  equivalent and they imply
each-other in a clear-cut fashion. It should be noted that all the rest
of the commutators (e.g. $[x, \psi] = 0, [\Pi_x, \psi] = 0, [x ,\bar\psi ] = 0 $) are zero. These, in turn, imply
that $[a, b] =0, [a^\dagger, b ] = 0, [a, b^\dagger ] = 0, [a^\dagger, b^\dagger ] = 0$, etc.

Against the backdrop of the above arguments, we have the following explicit (anti)commutators 
\begin{eqnarray}
&& [x(t),\, x(t)] = [\Pi_{x},\, \Pi_{x}] = 0 \;\Rightarrow [a,\, a] = 0, 
\; [a^{\dagger},\, a^{\dagger}] = 0,   \nonumber\\ 
&& [x(t),\Pi_{x}(t)] = [ x(t) ,\dot{x}(t)] = i \, \;\;\Rightarrow  [a,\, a^{\dagger}] = 1,  \nonumber\\
&& \{\psi(t),\, \Pi_{\psi}(t) \} \equiv \{\psi(t),\, \bar{\psi}(t)\}= - 1 \Rightarrow  
\{ b,\, b^{\dagger} \} = - 1,  \nonumber\\
&& \{\psi(t),\, \psi(t)\} = \{\bar{\psi}(t),\, \bar{\psi}(t)\} = 0 \nonumber\\
&&\Rightarrow \{b,\, b\} =  \{b^{\dagger},\, b^{\dagger}\} = 0,
\end{eqnarray}
which are the basic brackets of our present theory in terms of the creation and annihilation operators. 
All the other possible commutators (e.g. $[a, b] = 0, [a^\dagger, b^\dagger ] = 0,$ etc.), are 
zero in our theory. In our forthcoming   sections, we shall derive these  brackets from the symmetry properties  without using the  
definition of  canonical conjugate momenta ($\Pi_{x} = p_{x} = \dot {x}$ and $\Pi_{\psi} = {\partial L}/{\partial \dot{\psi}} = -i \,\bar{\psi}$). 
It may be worthwhile to mention, once again, that the latter definitions are {\it purely} mathematical in nature 
and there is almost {\it no} physical intuition involved in it.

We end this section with the remark that we have used here the definition of the canonical conjugate momenta
and spin-statistic theorem to obtain the basic brackets at the variable level (cf. (9)). When we express these brackets 
in terms of the mode expansions, we end up with the basic (anti)commutators amongst the creation and annihilation
operators (cf. (10), (11)). Thus, the (anti)commutators at the variable level are equivalent\footnote{The brackets, at the level of
creation/annihilation operators, are superior in the sense that vacuum state of the theory is defined in terms of annihilation operators and the particle
interpretation of quantum theory becomes quite transparent in this set-up.} to the (anti)commutators at the level
of creation/annihilation operators. We note that, in our present derivation, there has been {\it no} need of normal-ordering
at any stage. However, this idea becomes essential and important when we deal with the Hamiltonian formalism and express it 
(and other relevant quantities) in terms of the creation/annihilation
operators.

\section{Basic brackets from the {\it first} transformation  of the two $\mathcal{N} = 2$ 
SUSY symmetries: symmetry principles}

To derive the basic canonical  brackets amongst the creation and  annihilation operators, from the symmetry 
transformations ($ s_{1} $), first of all, we note that\footnote{We would like to lay emphasis on the fact that
the concept of continuous symmetries and generators in (12) is very basic even in the realm of classical mechanics where the (graded)Poisson
brackets are defined between two dynamical variables in the momentum phase space for a given physical system.}:
\begin{eqnarray}
 s_{1}\, \Phi =  - i\, [\Phi,\, Q ]_{\pm}, \qquad\qquad  \Phi = x,\, \psi,\, \bar{\psi}, 
\end{eqnarray}
where the $\pm $ signs, as  subscripts on the square bracket, denote the (anti)commutator for the generic 
variable  $\Phi$ being (fermionic)bosonic in  nature.  Here we have  already used the concept of spin-statistics theorem. 
It can be explicitly checked that the conserved charge $ Q = (\dot {x} + i \,\omega \,x)\, \psi $ can be 
expressed  in terms of  creation and annihilation operators, using the mode expressions (8), as 
\begin{eqnarray}
Q = \frac{+2 \,i \,\omega}{\sqrt{2\,\omega}}\, a^{\dagger}\, b,
\end{eqnarray}
where we have used $\dot{x} = (- i\, \omega/\sqrt{2 \,\omega})\,( a \,e^{- i \,\omega \, t}
 - a^{\dagger}\, e^{i \,\omega\,  t})$. The above charge automatically appears in the normal-ordered  form.
Thus, there is no need of the application of normal-ordering. This far, we have used two ideas of the
standard method of quantization. These are the spin-statistics theorem and normal-ordering.

We discuss here the explicit derivation of the basic (anti)commutators from the symmetry considerations. 
In particular, we express the symmetry (and the principles involved in these symmetry transformations) 
in the language of generators. To begin with, first of all, we focus on the following transformation
\begin{eqnarray}
 s_{1}x = - i\, [x,\,Q] = \psi.
\end{eqnarray}
Using the normal mode expression (8) and the expression for $ Q $ from (13), we obtain the following 
basic canonical commutators:
\begin{eqnarray}
[a,\, b] = [a^{\dagger},\, b] = [a^{\dagger},\, a^{\dagger}] = 0,\qquad  [a,\, a^{\dagger}] = 1
\end{eqnarray}
where we have compared the coefficients\footnote{Note that we are allowed to do this because $e^{-i\omega t}$ and $e^{-i\omega t}$
are the linearly independent solutions of the differential equation for the 1D SUSY-HO: $(\frac{d^2}{dt^2} + \omega^2)\, \Phi = 0$ 
where $\Phi = x, \psi, \bar\psi$.}  of $ e^{- i \,\omega\, t} $ and $e^{+ i \,\omega\, t}$ 
from the l.h.s. and r.h.s. Obviously,  the r.h.s. (i.e $ \psi = b\, e^{- i\, \omega\, t}$) contains only 
$ e^{- i \,\omega\, t} $ but the l.h.s.  contains both the  exponentials  $ e^{- i\, \omega\, t} $ 
as well as $ e^{ +i\, \omega\, t} $. Thus, it is clear that the coefficient of $ e^{ +i\, \omega\, t}$,
from the l.h.s., should be zero. Some of the basic brackets of (15) have been derived from this input. 
Moreover, we have also used the basic tricks of the (anti)commutators involving composite operators
(so that $[a, a^{\dagger}\, b] = [a, a^{\dagger}]\, b + a^{\dagger}\,[a, b],$ etc.).

Now, we concentrate on the transformations $ s_{1} \psi = 0 $ and $ s_{1}\, \bar{\psi} 
= i\,(\dot{x}+i\,\omega \,x)$. These can be written in terms of the generator $Q$ as
\begin{eqnarray}
&& s_{1}\, \psi = -i\, \{\psi,\,Q\} = 0, \nonumber\\  && s_1\, \bar{\psi} = 
-i\, \{\bar\psi,\,Q\} = i \,(\dot{x} + i\, \omega \,x).
\end{eqnarray} 
Plugging in the mode expansions from (8) and expression for $ Q $ from (13), we obtain the following 
basic brackets (i.e. (anti)commutators) 
\begin{eqnarray}
[a^{\dagger},\, b ] = 0, \qquad \qquad \{b,\, b \} = 0, 
\end{eqnarray} 
from the transformation $ s_{1}\,\psi = 0$. The other transformation $ s_{1}\,\bar{\psi} 
= - i\, \{\bar{\psi},\, Q \}$ leads to the following basic  brackets (i.e. a commutator and an anticommutator), namely;
\begin{eqnarray}
 [a^{\dagger},\, b^{\dagger}] = 0 ,\qquad \{b, \, b^{\dagger}\} = -1.
\end{eqnarray}
Ultimately,  we note that we have derived  the following  basic brackets from
the transformations $s_1$ on the generic variable $\Phi$
(i.e.  $ s_{1} \Phi = - i[\Phi,\, Q ]_{\pm} $ for $ \Phi = x,\psi ,\bar{\psi} $):
\begin{eqnarray}
&& [a,\, b] = 0, \quad [a^{\dagger}, \, b] = 0, \quad 
[a^{\dagger},\, a^{\dagger}] = 0,  \quad  \{ b,\, b\} = 0, \nonumber\\  &&
[a^{\dagger}, \, b^{\dagger}] = 0, \qquad \{ b, \,  b^{\dagger} \} =- 1,\qquad [a, \, a^{\dagger}] = +1, 
\end{eqnarray}
  which are seven in number. Out  of the above basic brackets, the non-vanishing  brackets are merely two
 (i.e  $\{b,\,  b^{\dagger}\} = -1,\;[a,\, a^{\dagger}]= +1$). We also point out that the use of Eq. (12) leads
 to the derivation of all possible brackets. We note that the symmetry considerations in (14) and (16) do {\it not}
 yield the bracket $\{b^{\dagger}, b^{\dagger}\} = 0$ which is required for the precise and complete quantization.

\section{Basic brackets from the {\it other} transformation of the  two $\mathcal{N} = 2$ 
SUSY symmetries: symmetry principles}

We dwell  a bit on the nilpotent transformations $s_{2}$ and concentrate on the transformation
of the bosonic variable $x$ as:
\begin{eqnarray}
 s_{2} \, x = - i\, [x,\, \bar{Q}]= \bar{\psi},
\end{eqnarray}
where the conserved charge $ \bar{Q}=\bar{\psi} \,(\dot{x} - i\, \omega \,x)$ can be expressed  in terms of the
mode expansions in (8) as:
\begin{equation}
 \bar{Q} = \frac{-2 \,i \,\omega}{\sqrt{2\,\omega}}\; b^{\dagger}\, a.
\end{equation}
The substitution   the mode expansions of (8) into (20)
leads to the emergence  of the following basic brackets from the transformations (20), namely;
\begin{equation}
[a,\, b^{\dagger}] = [a,\, a] = [a^{\dagger},\, b^{\dagger}] = 0,\qquad [a,\, a^{\dagger}] = 1,
\end{equation}
where we  have equated the coefficients of $ e^{-i\,\omega t}\, $ and $ e^{+i\,\omega\, t} $ 
from the l.h.s and r.h.s. Next, the trivial transformations 
$s_{2}\, \bar{\psi} = - i\,\{ \bar{\psi},\, \bar{Q} \} = 0$ yields the derivation of 
 $[a, \, b^{\dagger}] = 0$ and $\{b^{\dagger},\, b^{\dagger}\}= 0$ when we use the basic tricks of the
anticommutators with composite operators (e.g. $\{b^\dagger,\, b^\dagger\, a \} 
= \{b^\dagger,\, b^\dagger \}\, a - b^\dagger \, [b^\dagger,\, a ]$, etc.).
Finally, the transformations
\begin{eqnarray}
 s_{2}\,\psi = -i\, \{\psi,\, \bar{Q}\} = i\, (\dot{x}-i\,\omega\, x),
\end{eqnarray}
generates the basic brackets that are listed as follows:
\begin{equation}
 [a,\, b] = 0,\qquad \qquad \{b,\, b^{\dagger}\} = -1.
\end{equation}
 In the above derivation, we have compared the coefficients of $e^{ -\, i\,\omega\, t}$ and $e^{ +\, i\,\omega\, t}$
 from the l.h.s. and r.h.s. and used the basic tricks of the (anti)commutators with composite operators
(e.g. $\{ b, b^\dagger a \} = \{ b, b^\dagger \} a - b^\dagger [b, a]$).
Finally, we observe that $ s_{2}\Phi =- i\, [\Phi,\, \bar{Q}]_{\pm}$ (with $ \Phi = x,\,\psi,\,\bar{\psi}$) leads to the following basic (anti)commutation relations amongst the creation and annihilation operators:
\begin{eqnarray}
&& [a,\, b^{\dagger}] = [a,\, a] = [a^{\dagger},\, b^{\dagger}]= [a,\, b] = 0, \nonumber \\
&& \{b^{\dagger},\, b^{\dagger}\} = 0,\quad [a,\, a^{\dagger}] = + \,1, \quad \{b, \, b^{\dagger} \} = -\, 1.
\end{eqnarray}
A careful observation at (19) and (25) demonstrates  that we have already derived {\it all} the 
(non-)vanishing brackets amongst  the creation and annihilation operators 
 (i.e  $a,\, a^{\dagger},\, b,\, b^{\dagger}$).
 The non-vanishing brackets are $ [a,\, a^{\dagger}] = +1 $ and $ \{b,\, b^{\dagger}\} = -1 $
which are consistent with the ones derived in Sec. 3. We lay emphasis on the fact that the symmetry transformations ($s_2$)
do {\it not} produce the bracket $\{b,\, b\} = 0$ which is required for the complete  quantization.

\section{Basic brackets from the bosonic symmetry}

In Sec. 2, we have seen that the anticommutator of the nilpotent $\mathcal{N} = 2$ SUSY transformations 
produces a bosonic symmetry transformation ($s_{\omega}$). Under  this symmetry transformations, we have
the conserved  (i.e. ${\dot Q}_{\omega} = 0$) charge $Q_\omega =\frac{{p_x}^2}{2} + \frac{\,\omega^2\, x^2}{2} + \omega \, \bar\psi\, \psi$ that 
can be  expressed  in terms of the mode expansions (8) as
\begin{eqnarray}
 Q_{\omega}= H = \frac{\omega}{2}\, \left(a\,a^{\dagger}+\,a^{\dagger}\,a \right) + \omega \,b^{\dagger} \,b 
\equiv \,\omega\, \left (a^{\dagger}\,a + b^{\dagger}\,b \right ),
\end{eqnarray} 
where we have used the normal-ordering to make physical sense  out of $Q_{\omega}$. This expression would be used in the 
derivation of the transformations $s_{\omega}\,\Phi = \pm \, i\,[\Phi,\,Q_{\omega}]_{-}$
where $\Phi  =  x, \, \psi ,\bar{\psi}$ and $Q_\omega = H$ (that is given in Sec. 2 as well as in (26)).

To obtain the basic (anti)commutators, first of all, we focus on the transformation of the bosonic variable $x$. 
This can be written as
\begin{eqnarray}
s_{\omega}\,x=-i\,[x, \,Q_{\omega}] \, = \,\dot{x}.
\end{eqnarray}
The l.h.s.  and r.h.s.  of the above expression can be written in terms of the creation 
and annihilation operators and exponentials. The comparison of the coefficients of
 $e^{-i\,\omega \,t}$ and $e^{+\,i\,\omega \,t}$  (from the l.h.s. and r.h.s.) yields the following basic commutators:
\begin{eqnarray}
&& [a,\,b^{\dagger}]\, = \,[a, \,b]\,=\,[a, \,a]\,= [a^{\dagger}, \,b^{\dagger}] = 0 \nonumber\\
&& [a^{\dagger},\, b]\,=[a^{\dagger}, \,a^{\dagger}]\,=\,0, \qquad 
[a, \,a^{\dagger}]\,=\,1.
\end{eqnarray}
Thus, we note that the non-vanishing bracket is $[a, \,a^{\dagger}]\,=\,1$.
Let us now concentrate on the transformations:
\begin{eqnarray}
s_{\omega}\, \psi = i\,[\psi, \,Q_{\omega}]\,= \,\dot{\psi}, \qquad 
s_{\omega}\, \bar{\psi}\,=\,i\,[\bar{\psi}, \, Q_{\omega}]\,= \,\dot{\bar\psi}.
\end{eqnarray}
Plugging in the mode expansions (8) and comparing the coefficients  of
 $e^{-\,i\,\omega \,t}$ and $e^{+\,i\,\omega \, t}$ from l.h.s and r.h.s, we obtain the following
\begin{eqnarray}
&&[a, \, b]\,=\,[a^{\dagger}, \,b]\,=\,\{b,\, b\} = 0, \qquad\  \{b,\,b^{\dagger}\} \,=\,-1, \nonumber\\ &&
[a, \,b^{\dagger}] = [a^{\dagger}, \,b^{\dagger}] =\{b^{\dagger}, \,b^{\dagger}\}= 0,
\;\; \{b,\, b^{\dagger}\}\,= \,-1,
\end{eqnarray}
from the transformations $s_{\omega}\,\psi = \dot{\psi}$ and $s_{\omega}\,\bar{\psi} = \dot{\bar\psi}$, respectively.
Thus, we have obtained {\it all} the basic (anti)commutators of our theory where the non-vanishing brackets are 
$[a,\, a^{\dagger}] = 1$ and  $\{b, \,b^{\dagger}\} = -1$ (which are equivalent to the basic canonical 
brackets $[\,x,\,\Pi_{x}\,]\,=\,i,\;\;$ $\{\,\psi,\,\bar{\psi}\,\}\,=\,-1$ at the level of the variables).

\section{Conclusions}

In our present investigation, we have established that, for the models of the Hodge theory, 
the canonical quantization conditions can be achieved by exploiting  the spin-statistics theorem, 
normal-ordering and  symmetry principles. All these ideas are very nicely
backed (and bolstered  by the physical arguments and insights). For these  models, the mathematical definition of the canonical 
conjugate momenta, corresponding to the dynamical variables, are not required. 
We have corroborated the above statements 
in the case of 1D SUSY-HO where we have derived the basic (anti)commutators
amongst the creation/annihilation operators by exploiting the virtues of symmetry principles 
and have not used the mathematical definition of the canonical conjugate momenta (corresponding to the dynamical variables).
For the sake of comparison, we have also derived these brackets from the standard canonical 
quantization scheme (cf. Sec. 3)  so that the sanctity of our results could be clearly and firmly established
(in the case of 1D SUSY-HO).

Earlier attempts are present in literature where the alternative methods of the derivation of basic brackets
have been proposed. For instance, by exploiting the global spacetime Poincar{\'e} group and its generators, 
it has been shown (in the standard book on quantum field theory [10]) that the canonical brackets can be derived 
for the bosonic fields and  their creation/annihilation operators. However, 
in our present investigation and earlier works [6,7], we have exploited only the {\it internal}
symmetries of a given theory. Similarly, in a very nice piece of work by Wigner [11], the Heisenberg equations of motion 
have led to the derivation of basic canonical brackets where, once again, only the bosonic variables/fields
have been taken into account. Unlike our present work which is based on the internal symmetries 
and corresponding symmetry principles,
this attempt [11] is also {\it not} based on such type of internal symmetry considerations.

We lay emphasis on the observation that the ${\mathcal N} = 2$ SUSY transformations $s_1$ and $s_2$ (and 
their generators $Q$ and $\bar Q$) do {\it not} produce {\it all} the (anti)commutators of the theory.
As pointed out after (19) and (25), the brackets $\{b^{\dagger}, b^{\dagger} \} = 0$ and $\{b, b \} = 0$ are {\it not}
produced by the pairs $(s_1,\, Q)$ and $(s_2,\, \bar Q)$, respectively. However, the transformations $s_{\omega}$ 
(generated by $Q_{\omega}$) produce all the appropriate (anti)commutators as is illustrated in (30).
Thus, we have observed that the results, produced by $Q$ and $\bar Q$ {\it together}, 
emerge automatically by using $Q_{\omega} = H$.
There is a deeper mathematical reason behind it. 
It has been shown in [1] that the set ($Q, \, \bar Q,\, Q_{\omega} \equiv H$)
provides the physical realizations of the de Rham cohomological operators
of differential geometry where 
 $ Q_{\omega} \equiv H$
is identified with the Laplacian operator which is equal to the anticommutator of the exterior and co-exterior derivatives. 
The (co-)exterior derivatives are identified with the $Q$ and $\bar Q$ in the language of the symmetry generators. Thus, it is clear
that the consequences, that emerge from $Q_{\omega}$, would be equivalent to the results obtained by $Q$ and $\bar Q$ 
separately and independently (see, e.g. [1], [16] for detailed discussions). It is worthwhile to mention
here that on a
compact manifold without a boundary, we define a set of {\it three} operators $(d, \delta, \Delta)$
which are known  as exterior derivative, co-exterior derivative and Laplacian
operators, respectively. They obey the Hodge algebra: $d^2 = \delta^2 =0, \Delta = \{d, \, \delta\} 
\equiv  (d + \delta)^2, \, [\Delta, \, d] = [\Delta, \, \delta ] =0$ which shows that $\Delta$ is like 
a Casimir operator (see, e.g. [12-15]).

We have proven that the 1D model of a rigid rotor [17], ${\mathcal N} = 2$ SUSY quantum mechanical model with any 
arbitrary superpotential [16],  $\mathcal{N} = 2$ SUSY model for the motion of a charged particle
under influence of a magnetic field [18], free 4D Abelian 2-form and 6D Abelian 3-form gauge theories [19-21], etc., 
are models for the Hodge theory. For all these models, we can perform the canonical quantization without taking 
recourse  to the mathematical definition of the canonical conjugate momenta. 
It would be very nice future endeavor for us to accomplish these goals in a clear-cut fashion. 
These are the problems we are intensively involved with,  at present, and 
the results and findings would be reported in our future publications [22]. \\

\vskip 0.5cm

\noindent
{\bf Acknowledgements} \\

\noindent
One of us (AS) would like to gratefully acknowledge the financial support
from  CSIR, Government of India, New Delhi, under its SRF-scheme. All of us
would like to thank our colleague Mr. S. Krishna for carefully reading the manuscript
as well as for offering us some crucial, decisive and valuable inputs on the topic of our present investigation.

\end{document}